\documentclass[twocolumn,superscriptaddress,aps,preprintnumbers,amsmath,amssymb,,sort&compress,nofootinbib
]{revtex4-1}

\usepackage{amsfonts}
\usepackage{ascmac}
\usepackage{bm}
\usepackage{comment}
\usepackage{ifpdf}
\usepackage{slashed}
\usepackage{color}
\usepackage[mathscr]{eucal}
\usepackage[utopia]{mathdesign}

\usepackage{cancel}

\ifpdf
  \usepackage{graphicx}     
  \usepackage[bookmarksopen,colorlinks=true,linkcolor=bblue,citecolor=bblue,urlcolor=ppink]{hyperref}
\else     
  \usepackage[dvipdfmx]{graphicx}     
  \usepackage[dvipdfmx,bookmarksopen,colorlinks=true,linkcolor=bblue,citecolor=ppink,urlcolor=darkred]{hyperref}
\fi

\definecolor{red}{rgb}{1,0,0}
\definecolor{darkred}{rgb}{0.6,0,0}
\definecolor{darkgreen}{rgb}{0.992447,0.623778,0.034597}
\definecolor{ppink}{rgb}{1,0.4,0.4}
\definecolor{bblue}{rgb}{0.284602,0.317763,0.963947}

\newcommand{\vev}[1]{ \left< {#1} \right> }

\newcommand{\prn}[1]{\left( {#1} \right)}
\newcommand{\com}[1]{\left[ {#1} \right]}

\newcommand{\dd}{\mathrm{d}}
\newcommand{\Mpl}{M_{\rm Pl}}

\newcommand{\abs}[1]{\left\vert {#1} \right\vert}

\def\Mpl{M_{\rm pl}}

\begin{document}


\title{
Simple Cosmological Solution to the Higgs Instability Problem in Chaotic Inflation\\[.5em]
and Formation of Primordial Black Holes
}

\author{Masahiro Kawasaki}
\affiliation{ICRR, University of Tokyo, Kashiwa, 277-8582, Japan}
\affiliation{Kavli IPMU (WPI), UTIAS, University of Tokyo, Kashiwa, 277-8583, Japan}
\author{Kyohei Mukaida}
\affiliation{Kavli IPMU (WPI), UTIAS, University of Tokyo, Kashiwa, 277-8583, Japan}
\author{Tsutomu T.~Yanagida}
\affiliation{Kavli IPMU (WPI), UTIAS, University of Tokyo, Kashiwa, 277-8583, Japan}

\begin{abstract}
\noindent
We revisit the compatibility between \textit{chaotic inflation}, 
which provides a natural solution to the initial condition problem,
and the \textit{metastable electroweak vacuum}, 
which is suggested by the results of the LHC
and the current mass measurements of top quark and Higgs boson.
It is known that chaotic inflation poses a threat to the stability of the electroweak vacuum
because it easily generates large Higgs fluctuations during inflation 
or preheating and triggers the catastrophic vacuum decay.
In this paper, we propose a simple cosmological solution in which 
the vacuum is stabilized during chaotic inflation, preheating and afterwards.
This simple solution naturally predicts the formation of primordial black holes. 
We find interesting parameter regions where the present dark matter density is provided by them.
Also, the thermal leptogenesis can be accommodated in our scenario.
\end{abstract}

\date{\today}
\maketitle
\preprint{IPMU 16-0072}

\section{Introduction}\label{sec: intro}
Inflation has now become an essential part of the standard cosmology.
It elegantly solves major problems of Big Bang cosmology,
namely the horizon/flatness problems,
and simultaneously provides the origin of 
primordial density fluctuations~\cite{Mukhanov:1981xt,Hawking:1982cz,Starobinsky:1982ee,Guth:1982ec,Bardeen:1983qw}
as observed by the Planck satellite~\cite{Ade:2015xua}.
The idea of chaotic inflation~\cite{Linde:1983gd} is particularly attractive
among many mechanisms proposed so far.
This is because it persists under large quantum fluctuations 
at the very beginning of the Universe, \textit{i.e.}~Planck time.
Therefore, the initial condition problem is naturally solved. 
In addition, as a result of the trans-Planckian field excursion,
large primordial tensor perturbations are predicted~\cite{Starobinsky:1980te,Lyth:1996im},
which makes this paradigm testable in the forthcoming future.
The current upper bound on the tensor-to-scalar ratio of $r < 0.07$ ($95 \%$ C.L.)
is given by the combined analysis of  Planck and BICEP2/Keck Array~\cite{Array:2015xqh}.

The results of LHC experiments strongly support that the Standard Model (SM) 
would be valid above the electroweak scale, contrary to the naturalness argument. 
It is tempting to think boldly that
there is no new physics beyond the SM up to very high energy scales,
say the Planck scale.
Interestingly, the current measurements of top and Higgs masses suggest that our vacuum is likely to be metastable
if there are no new physics contributions to the Higgs potential~\cite{Sher:1988mj,Arnold:1989cb,Degrassi:2012ry,Buttazzo:2013uya,Bednyakov:2015sca}.
The lifetime of our vacuum exceeds the current age of the Universe~\cite{Espinosa:2007qp,Rose:2015lna},
and the energy scale, $h_\text{max}$, above which the four-point coupling becomes negative is around
$10^{10}$ GeV for the best-fit values of SM parameters~\cite{Buttazzo:2013uya,Bednyakov:2015sca}.

We assume that the Universe experiences the chaotic inflationary epoch at its very early stage,
taking the initial condition problem seriously.
Then, the metastability of our vacuum leads to an interesting consequence:
a stabilization mechanism, such as a positive Hubble-induced mass term, is required 
so as to save the vacuum 
during inflation~\cite{Espinosa:2007qp,Lebedev:2012sy,Hook:2014uia,Espinosa:2015qea}.
The Higgs-curvature coupling, $\xi R \hat H^\dag \hat H$, plays this role naturally,
since the Higgs universally couples to all the would-be inflatons,
regardless of the chaotic inflationary dynamics 
before the last one responsible for the observed CMB fluctuations.
Here $\xi$ is the non-minimal coupling, $R$ is the Ricci scalar and $\hat H$ is the Higgs doublet.
There is a lower bound on the non-minimal coupling,
$\xi \gtrsim \mathcal{O} (0.1)$, so that the Hubble-induced mass is effective during inflation~\cite{Espinosa:2007qp,Herranen:2014cua}.

After the inflation, however, this effective mass term oscillates between positive and negative values
owing to the coherent oscillation of inflaton.
As a result, large Higgs fluctuations are produced by the tachyonic resonance~\cite{Bassett:1997az,Tsujikawa:1999jh,Herranen:2015ima}, 
potentially triggering the vacuum decay.
One of the authors put an upper bound on the coupling, $\xi \lesssim \mathcal O (10)$,
above which the resonance immediately activates the vacuum decay~\cite{Ema:2016kpf}.\footnote{
	See also \cite{Kohri:2016wof}.
}
Moreover, even for the non-minimal coupling satisfying these bounds, say $\xi \sim \mathcal O (1)$,
production of large Higgs fluctuations is probable.
Although the oscillation-averaged mass term from $\xi R \hat H^\dag \hat H$ 
is still large enough right after the end of resonance in this case,
this mass term drops faster than the Higgs fluctuations
by the cosmic expansion afterwards, and eventually the vacuum decay could occur.

Let us rephrase this issue in a more intuitive way.
As already mentioned,
if the SM plasma remains close to thermal equilibrium,
it is known that the lifetime of the electroweak vacuum is longer than the current age of the Universe,
even with $T \sim \Mpl$ for the best-fit values of SM parameters~\cite{Espinosa:2007qp,Rose:2015lna}.
Hence, one possible answer may be to generate the hot Universe \textit{adiabatically} after chaotic inflation.
The first example is to take the non-minimal coupling small enough, for instance close to
the conformal coupling $\xi = 1/6$~\cite{Herranen:2015ima}.
Another one would be efficient production of electroweak gauge bosons and top quarks during thermalization.
In particular, the direct decay of the inflaton to SM particles, which is required for a complete reheating may be helpful.
Yet, this mechanism strongly depends on thermalization processes and further studies are required.

In this paper, we propose a simple cosmological solution
in which the vacuum is always stabilized during chaotic inflation, preheating, and afterwards
in a wide range of $\xi$, without modifying the Higgs potential. 
Our scenario consists of SM plus the curvature coupling and two inflations.
We assume that a new inflation~\cite{Albrecht:1982wi,Linde:1981mu} takes place 
after the end of the last chaotic inflation
which provides density perturbations observed today.
The initial condition problem of the new inflation
is solved dynamically during the chaotic inflation~\cite{Izawa:1997df}.
Note that the new inflation has to start
before the onset of the vacuum decay in the inflaton oscillation regime after the chaotic inflation.
Hence, its scale must be close to the chaotic one.
After the beginning of the new inflation, 
the Higgs quickly settles down to its potential origin again
thanks to the curvature coupling.
The crucial difference between new inflation and the chaotic one
is that, after inflation, the inflaton oscillates with an amplitude
that is much smaller than the Planck scale.
Hence, even the narrow resonance does not take place after the new inflation. 
Although the gravitational Higgs production~\cite{Ford:1986sy} right after the new inflation is 
inevitable~\cite{Herranen:2015ima},
we can avoid this constraint for a suitable choice of the new inflation scale.

Moreover, the new inflation can produce primordial black holes (PBHs)
because it is free from the COBE normalization in our scenario~\cite{Kawasaki:1997ju,Kawasaki:1998vx,Kanazawa:2000ea,Frampton:2010sw}.
A sizable amount of PBHs can be produced in a wide range of PBH mass,
or only within a narrow range, depending on model parameters.
We demonstrate that PBHs can be a dominant component of dark matter (DM) in both cases.
Also, the reheating temperature tends to be high because
the new inflation scale has to be close enough to avoid the decay of the electroweak vacuum.
The predicted value is consistent with the thermal leptogenesis~\cite{Fukugita:1986hr}
in most of the parameter regions.
Therefore, our model can be regarded as one of minimal scenarios
consistent with the idea of chaotic inflation and the metastable electroweak vacuum.

\section{Chaotic Inflation and Metastable Vacuum}
\label{sec: metastability}

First of all, let us briefly recall 
the motivation of chaotic inflation models
and the reason for a Planckian field value.
As emphasized in the Introduction,
chaotic inflation is special,
for it can take place under natural initial conditions:
$\dot \phi^2 \sim (\nabla \phi)^2 \sim V(\phi) \sim \Mpl^4$, 
$l \sim H^{-1} \sim \Mpl$, where $ \phi$ is a scalar field,
$l$ is the size of the Universe and $H$ is the Hubble parameter.
As a result, it can occur even in a \textit{closed} Universe, 
which otherwise would collapse within $\mathcal O (\Mpl^{-1})$.
In order for chaotic inflation to take place, 
its potential should satisfy the slow roll condition.
It may be characterized by the smallness of potential slow roll parameters
$\epsilon_V, | \eta_V | \ll 1$ where
$\epsilon_V \equiv (\Mpl^2/2) (V'/V)^2$ and $\eta_V \equiv \Mpl^2 V''/V$.
As an illustration, suppose that the potential is dominated by $V \sim \lambda_n \phi^n / n$.
Interestingly, the initial condition $V \sim \Mpl^4$ indicates 
that a typical initial field value is much larger than the Planck scale
for $ \lambda_n \ll 1$ (in Planck unit) which is suggested by the slow roll conditions.
In addition, one can see that the slow roll condition is violated typically at $\phi_\text{end} \sim \Mpl$.
Therefore, chaotic inflation begins with $ \phi \gg \Mpl$, and ends at $\phi \sim \Mpl$.
After that, the inflaton starts to oscillate with a Planckian amplitude,
which could trigger the vacuum decay by producing large Higgs fluctuations.

Next, we summarize basic properties of the metastable electroweak vacuum in the chaotic inflation paradigm.
In a quasi de Sitter Universe during inflation, light fields whose mass is smaller than the Hubble parameter $H$
develop fluctuations with an amplitude of $H/2\pi$,
which corresponds to  the Gibbons-Hawking temperature~\cite{Gibbons:1977mu}.
The field value of Higgs, $h_\text{max}$, above which the Higgs four-point coupling turns into negative,
is around $10^{10}$ GeV for the best-fit values of SM parameters,
although it strongly depends on the top quark mass~\cite{Buttazzo:2013uya,Bednyakov:2015sca}.
In the following discussion, we take the center value as a reference unless otherwise stated.
Since a typical value of the Hubble parameter during chaotic inflation, $H_\text{ch} \sim 10^{14}$ GeV,
is much larger than this scale, long-wavelength modes of Higgs easily climb over the potential barrier.
Namely, our vacuum decays via the Hawking Moss instanton~\cite{Hawking:1981fz}.
Thus, there is a severe tension between chaotic inflation and
the Higgs metastability.

The curvature coupling, $\xi R \hat H^\dag \hat H$, provides a natural solution
because the Higgs field couples with would-be inflatons universally.
Here $\xi$ is the non-minimal coupling, $R$ is the Ricci scalar and $\hat H$ is the Higgs doublet.
As a result, 
it yields a sizable Hubble-induced mass term regardless of the chaotic inflationary dynamics,
until the end of the last one responsible for density perturbations observed today.
During inflation, it induces an effective mass squared of $12 \xi H^2$.
Hence, our vacuum is stabilized against fluctuations
for the non-minimal coupling satisfying 
$\xi \gtrsim \mathcal O (0.1)$~\cite{Espinosa:2007qp,Herranen:2014cua}.

However, as already mentioned in the beginning of this section,
the curvature coupling oscillates violently after inflation 
due to the coherent oscillation of the inflaton with a Planckian amplitude.
As a result, the Higgs acquires large fluctuations in the preheating stage,
which poses a threat to the stability of our vacuum again.
In the rest of this section, we briefly repeat the main results given in Ref.~\cite{Ema:2016kpf}.

The Einstein equation relates the Ricci curvature with a trace of the energy momentum tensor.
In the inflaton-dominated era, the curvature can be expressed as
\begin{align}
	R = \frac{1}{\Mpl^2} \com{ 4 V(\phi) - \dot \phi^2 },
\end{align}
where $\phi$ is a real scalar field which causes chaotic inflation and $V(\phi)$ is its potential.
One can see that the effective mass term oscillates between positive and negative values.
Therefore, the tachyonic resonance takes place~\cite{Bassett:1997az,Tsujikawa:1999jh,Herranen:2015ima}.

We adopt the following mode expansion of the Higgs field:
\begin{align}
	h (x) = \int \frac{\dd^3 k}{\com{2 \pi a (t)}^{3/2}}
	\com{ \hat a_k h_k (t) e^{i \bm{k} \cdot \bm{x}} + \text{H.c.}},
\end{align}
where we denote one component of the Higgs field as $h$,
$\bm{k}$ is a comoving momentum, $a(t)$ is the scale factor,
$h_k$ is a mode function of the Higgs field,
and $\hat a_{\bm{k}}^{(\dag)}$ is the annihilation (creation) operator 
satisfying $[ \hat a_{\bm{k}}, \hat a_{\bm{k'}}^\dag ] = \delta (\bm{k} - \bm{k'})$
and $[ \hat a_{\bm{k}}, \hat a_{\bm{k'}} ] = [ \hat a_{\bm{k}}^\dag, \hat a_{\bm{k'}}^\dag ] = 0$.
The normalization of the Wronskian condition is taken to be
$h_{\bm{k}} \dot h_{\bm{k}}^\ast - h_{\bm{k}}^\ast \dot h_{\bm{k}} = i$.
Ignoring interactions with SM fields,
one obtains the following mode equation:
\begin{align}
	0 = \com{ \frac{\dd^2}{\dd t^2} + \frac{\bm{k}^2}{a^2 (t)}
	+ \prn{ \xi  - \frac{1}{6} }R - \frac{1}{2} \prn{ \dot H + \frac{H^2}{2}} } h_k (t).
	\label{eq:mode_general}
\end{align}
For simplicity, suppose that the inflaton oscillates 
under a quadratic potential after inflation.\footnote{
	For other powers of  inflaton potential, 
	we may apply the following discussion
	by replacing $m_\phi$ to $\sqrt{V'/\phi}_{\dot \phi \sim 0}$
	for an order of magnitude estimation.
} 
Then the inflaton oscillation can be approximated by 
$\phi (t) \simeq \Phi (t) \cos ( m_\phi t )$
where $\Phi$ is the oscillation amplitude decreasing proportional to 
$a^{-3/2}\propto t^{-1}$,
and $m_\phi$ is the inflaton mass.
As mentioned at the beginning of this section, its initial amplitude is Planckian $\Phi_\text{ini} \sim \Mpl$.
Plugging this approximated solution into Eq.~\eqref{eq:mode_general}
and treating the inflaton oscillation as a background,
we arrive at the familiar Mathieu equation:
\begin{align}
	0 = \com{ \frac{\dd^2}{\dd (m_\phi t)^2} + A_k (t) - 2q (t) \cos (2 m_\phi t) } h_k (t),
	\label{eq:mathieu}
\end{align}
where 
\begin{align}
	A_k (t) &\equiv \frac{\bm{k}^2}{a^2 (t) m_\phi^2} + \frac{\xi \Phi^2 (t)}{2 \Mpl^2}, &
	q (t) & \equiv - \frac{3}{4} \prn{ \xi - \frac{1}{4} } \frac{\Phi^2 (t)}{\Mpl^2}.
	\label{eq:mathieu_prm}
\end{align}
If the resonance parameter is larger than unity, $|q| \gtrsim 1$,
the tachyonic resonance produces large Higgs fluctuations
below a typical momentum, 
$k/a \lesssim p_\ast \equiv |q|^{1/4} m_\phi / \sqrt{x}$
with $x \simeq 0.85$~\cite{Dufaux:2006ee}.
Since the growth rate of Higgs fluctuations is rather power-law-like in this case,
first few oscillations of inflaton would be important.
Higgs fluctuations after the $j$th passage of $\phi \sim 0$ can be evaluated as
\begin{align}
	\vev{h^2} (t_j) &\sim 
	\frac{1}{16 \pi^2} \sqrt{\frac{\pi}{2}} \frac{a^2 (t_\text{ini})}{a^2 (t_j)} p_\ast^2 (t_\text{ini}) 
	e^{n (t_j) \mu \sqrt{\xi} \frac{\Phi_\text{ini}}{\Mpl}},
	\label{eq:higgs_fluc}
\end{align}
where $\Phi_\text{ini}$ is an initial amplitude of the inflaton at the onset of oscillation,
a numerical constant is estimated as $\mu \simeq 2$,
and 
$n (t_j) \equiv \sum_{i=1}^j \Phi(t_i)/\Phi_\text{ini}$ denotes small time dependence 
growing logarithmically in time.

Long-wavelength modes of Higgs acquire a tachyonic effective mass,
if the Higgs fluctuations exceed the effective mass term from the curvature coupling,\footnote{
	If the inflaton instantaneously decays into radiation in a dark sector that is sequestered from SM,
	the effective mass term from the curvature coupling vanishes,
	and also Higgs does not acquire the thermal mass.
	In this case, the vacuum decay is almost inevitable unless it is close to the conformal coupling $\xi = 1/6$.
}
\textit{i.e.}~$3 |\tilde \lambda | \langle h^2 \rangle \gtrsim 
m_\text{eff}^2 \equiv \xi \overline{R} \sim |q| m_\phi^2$
with the over-line being an oscillating average and $\tilde \lambda \sim - \mathcal O (10^{-2})$ being
the negative four-point coupling of Higgs.\footnote{
	The infrared scale of our system is at least the Hubble parameter
	which is much larger than $h_\text{max}$.
	Therefore, we can use the running coupling constant above $10^{10}$ GeV
	that is estimated as $\tilde \lambda \sim - \mathcal O (10^{-2})$~\cite{Buttazzo:2013uya,Bednyakov:2015sca}.
}
For a large-enough non-minimal coupling $\xi$, this inequality is fulfilled.
Then, the Higgs field falls into the unwanted deeper minimum due to the tachyonic effective mass
before the end of the resonance.
If the vacuum survives the preheating stage,
the resonance ends at $|q| \lesssim 1$.\footnote{
 	Note that the back-reaction from the non-minimal coupling $\xi$ is irrelevant in our case.
	This is because, typically, the Higgs four-point coupling is much larger.
	The above inequality for the vacuum decay indicates that
	the backreaction/rescattering from the Higgs four-point coupling trigger
	the catastrophic decay.
}
This consideration gives the following upper bound on $\xi$
so that the vacuum does not decay during the resonance~\cite{Ema:2016kpf}\footnote{
	The bound depends on $\tilde \lambda$ only logarithmically,
	and thus its precise absolute value is not important 
	while its sign is crucial.
	Also, this bound was confirmed by a classical lattice simulation~\cite{Ema:2016kpf}.
}
\begin{align}
	\xi \lesssim 
	20 \times \prn{\frac{2}{n \mu}}^2  \prn{\frac{\Mpl}{\Phi_\text{ini}}}^2.
	\label{eq:upperbound_xi}
\end{align}
If this condition is violated, the Higgs fluctuations immediately overcome the effective mass
$m_\text{eff}^2 = \xi \overline{R}$
for a few oscillations of inflaton, $m_\phi t_\text{dec} \sim 10$.

If the non-minimal coupling lies in the following range, $\mathcal O (0.1) \lesssim \xi \lesssim \mathcal O (10)$,
our vacuum survives against fluctuations during inflation and the resonance.
Still, as can be seen from Eq.~\eqref{eq:higgs_fluc},
Higgs acquires large fluctuations comparable to or larger than the Hubble scale,
though the oscillation-averaged mass term, $m_\text{eff}^2 = \xi \overline{R}$, stabilizes
the Higgs field at the electroweak vacuum right after the end of the resonance in this case.
The dynamics of the Higgs after the resonance could be complicated,
since the interactions with other SM particles become relevant at that time scale,
contrary to the very short time scale of vacuum decay at the preheating stage, $m_\phi t_\text{dec} \sim 10$.
As an illustration, let us neglect these effects and see what would be expected.
Afterwards ($|q| \lesssim 1$),
the Higgs fluctuations decrease by the cosmic expansion, $\langle h^2 \rangle \propto a^{-2}$.
The fluctuations become smaller than $h_\text{max} \sim 10^{10}$ GeV,
at $(m_\phi t_\text{stb}) \sim 10^4 (\xi/2)^{3/8} (\Mpl / \Phi_\text{ini})^{1/4} e^{(3/\sqrt{2})[ \sqrt{\xi/2} \, (n_\text{end}\mu / 2) (\Phi_\text{ini}/\Mpl )  - 1]}$ where $n_\text{end}$ is evaluated at the end of resonance.
Here we denote the time after which the Higgs fluctuations become smaller than $h_\text{max}$ as $t_\text{stb}$.
However, the effective mass term from the curvature coupling decreases faster,
$m_\text{eff}^2 = \xi \overline{R} \propto a^{-3}$.
And thus, $3 |\tilde \lambda | \langle h^2 \rangle$  
can exceed $m_\text{eff}^2 = \xi \overline{R}$ at~\cite{Ema:2016kpf}
\begin{align}
	( m_\phi t_\text{dec} ) \sim 9\times 10^3 \prn{\frac{\xi}{2} }^\frac{3}{4} \prn{ \frac{\Phi_\text{ini}}{\Mpl} }^\frac{1}{2}
	e^{-3 \sqrt{2}\, \prn{ \sqrt{\frac{\xi}{2}} \frac{n_\text{end} \mu}{2}\frac{\Phi_\text{ini}}{\Mpl} - 1}}.
	\label{eq:upper_xi}
\end{align}
Comparing $t_\text{dec}$ with $t_\text{stb}$,
one finds that the Higgs field may escape from the electroweak vacuum
even after the end of the resonance, 
for instance in the case of  $\xi \gtrsim 1$ for $\Phi_\text{ini} \gtrsim \Mpl$.
Note again that production of other SM particles, like electroweak gauge bosons, is neglected.
See discussion below.

Several remarks are in order before presenting our model.
First, note that the above bounds should be regarded as an order of magnitude estimation,
and also, the bounds depend on a precise form of the chaotic inflation potential,
given that the efficiency of resonance exponentially depends on the initial amplitude
and the numerical constant $n_\text{end}\mu$.
To obtain precise bounds on $\xi$ and discuss its viable parameter space, 
we have to fix a model of chaotic inflation and solve the equations of motion numerically model by model.
Second, in deriving these bounds, we have neglected production of other SM particles.\footnote{
	Its possible effects are discussed qualitatively in Ref.~\cite{Ema:2016kpf}.
}
In particular, the decay of inflaton into SM particles which is required for a complete reheating could be important.
Since the thermal mass of the Higgs decreases slowly $m_\text{th}^2 \propto a^{-3/4}$,
it may stabilize the Higgs instead of $m_\text{eff}^2 = \xi \overline{R}$  after the preheating.\footnote{
	During the preheating, the decay of the electroweak vacuum takes place within a few oscillations of inflaton.
	Hence, we expect perturbative production of other SM particles, for instance via a dimension five Planck-suppressed decay of inflaton,
	may not play the role during the preheating~\cite{Ema:2016kpf}.
}
Yet, other SM particles also give potentially dangerous fluctuations to the Higgs, 
although it is known that 
the stabilization effect from the thermal mass is dominant for the observed SM parameters 
\textit {once the system is thermalized}.
Hence, this mechanism depends on thermalization processes,
and further studies are required.

In the rest of this paper,
we provide a simple cosmological solution 
where the vacuum is always stabilized during inflation, preheating, and afterwards
in a wide range of $\xi$,
regardless of details of thermalization processes and a chaotic inflation potential.

\section{New Inflation after Chaotic inflation}
\label{sec: double_inf}

Our model consists of SM with a sizable Higgs-curvature coupling, $\xi R \hat H^\dag \hat H$,
and two inflations:
one is a chaotic inflation providing the origin of density perturbations observed today,
and the other is a new inflation to suppress the resonance at the preheating stage.
As we will see, chaotic inflation dynamically solves the initial condition problem of the new inflation,
and thus it is interesting to assume that  
the pre-inflation is governed by the chaotic one~\cite{Izawa:1997df}.
We only assume that the chaotic inflaton, $\phi$, oscillates with a quadratic potential after inflation
but do not specify its precise form at its large field value responsible for chaotic inflation,
because it is irrelevant in the following discussion.
We consider the following potential for the new inflation:
\begin{align}
	{\cal V} (\varphi) &= 
		\prn{ v^2 - g \frac{\varphi^4}{\Mpl^{2}}}^2 - \kappa v^4 \frac{\varphi^2}{2\Mpl^2}  - \varepsilon v^4 \frac{\varphi}{\Mpl},
	\label{eq:potential_new}
\end{align}
where $\Mpl$ denotes the reduced Planck scale,
$v$ determines the scale of the new inflation and $g,\kappa$ and $\varepsilon$ are dimensionless couplings
whose typical sizes are specified later.
The first two terms respect a $\mathbb Z_2$-symmetry, $\varphi \to - \varphi$,
while the third term can be regarded as an order parameter of $\mathbb Z_2$-breaking,
which solves the initial condition and domain wall
problems of the new inflation as we see below.
We implicitly assume that the second term is somehow suppressed since otherwise
the new inflation does not take place.
In the following discussion, we take two bench marks for $\kappa$: 
$ \kappa \sim \varepsilon^2$ and  $\kappa \sim \mathcal O (0.1)$.\footnote{
	The value, $\kappa \sim \varepsilon^2$, means that  
	the coupling $\kappa$ is sufficiently small not to affect the dynamics of the new inflation.
	Also, one might expect $\kappa \sim \varepsilon^2$ at least because the third term completely breaks $\mathbb Z_2$,
	even if $\kappa$ is somehow suppressed at the beginning.
}
The predicted PBH spectra
significantly differ from each other as we will see.

To stabilize the new inflation potential during the chaotic one,
a positive Hubble-induced mass term is required.	For instance, the following interaction, 
$\mathcal L_\text{int} =  -c^2 \phi^2 \varphi^2 / 2$, suffices
with $c$ being a small coupling, $c \sim \mathcal O (10^{-5})$,
and $\phi$ being a real scalar field responsible for chaotic inflation.
The initial condition of the new inflation is dynamically determined~\cite{Izawa:1997df} 
by the balance between this term and the $\mathbb Z_2$-breaking term
\begin{align}
	\varphi_\text{ini} \simeq \prn{\frac{ \varepsilon v^4}{c^2 \overline{\phi^2} \Mpl}}_\text{ini}
	\sim v \prn{ \frac{\varepsilon \Mpl}{v} },
	\label{eq:new_ini}
\end{align}
where quantities with an over-line are averaged over oscillation period of $\phi$.
In the second similarity, we have used $c^2 \overline{\phi^2}|_\text{ini} \sim H^2|_\text{ini}
\sim v^4/M_\text{pl}^2$, which is expected at the onset of the new inflation.\footnote{
	Here, for simplicity, we have taken a coupling $c$ so that $c^2 \overline{\phi^2} |_\text{ini}
	\sim H^2 |_\text{ini}$.
}
We have assumed that the coupling $\kappa$ is smaller than unity, $ \kappa \ll 1$.
One can see that the $\mathbb Z_2$-breaking term should be suppressed
because of the condition $\varphi_\text{ini} < v$ [See also explanation below Eq.~\eqref{eq:end_newinf}].

Now we are in a position to discuss the inflationary dynamics of our model.
After chaotic inflation, the inflaton, $\phi$, oscillates around its potential minimum.
After a period, $t_\text{osc}$, 
the potential energy of $\varphi$ dominates the Universe and the new inflation begins.
We have the following relation between the new inflation scale, $v$, 
and the period, $t_\text{osc}$:
\begin{align}
	v \simeq 
	2 \times 10^{15} \, \text{GeV} \,
	\prn{ \frac{m_\phi}{10^{13}\, \text{GeV}} }^\frac{1}{2} \prn{ \frac{10}{m_\phi t_\text{osc}} }^\frac{1}{2},
	\label{eq:bound}
\end{align}
with $m_\phi$ being the mass of the chaotic inflation.
On the other hand, the period, $t_\text{osc}$, should be 
shorter than the time scale of vacuum decay, $t_\text{dec}$,
which depends on $\xi$ as estimated in the previous section:
\begin{align}
	t_\text{osc} < t_\text{dec} (\xi).
	\label{eq:xi_inequality}
\end{align}
For the $\xi$ dependence of $t_\text{dec}$,
see discussion around Eqs.~\eqref{eq:upperbound_xi} and \eqref{eq:upper_xi}.
Together with Eq.~\eqref{eq:bound}, this inequality puts a lower bound on
the new inflation scale, $v$, as a function of $\xi$.
If the bound is satisfied, the new inflation starts before the vacuum decay is triggered.
In the following discussion, we take $m_\phi t_\text{osc} \sim 10$ as a reference.
This value allows the non-minimal coupling as large as $\xi \sim \mathcal O (10)$.
Soon after the new inflation sets in, the Higgs field immediately moves back to its origin
because the sizable Hubble-induced mass term is generated 
by the potential energy of the new inflation, $ v^4$.

The new inflation lasts till its slow roll conditions are violated.
The potential slow roll parameters are given by
\begin{align}
	\epsilon_{\mathcal V} &\equiv \frac{\Mpl^2}{2} \prn{ \frac{\mathcal V'}{\mathcal V} }^2
	\simeq
	\frac{1}{2} \prn{ - \varepsilon - \kappa \frac{\varphi}{\Mpl}- 8 g \frac{\varphi^2}{v^2} \frac{\varphi}{\Mpl} + \cdots}^2, \\
	\eta_{\mathcal V} & \equiv \Mpl^2 \frac{\mathcal V''}{\mathcal V}
	\simeq - \kappa
	- 24 g \frac{\varphi^2}{v^2} +\cdots.
\end{align}
The slow roll regime ends at
\begin{align}
	\varphi_\text{end} \simeq \prn{ \frac{1}{24 g} }^\frac{1}{2} v.
	\label{eq:end_newinf}
\end{align}
Recalling Eq.~\eqref{eq:new_ini},
one can see that the $\mathbb Z_2$-breaking parameter should be so small,
$\varepsilon g^{1/2} \Mpl /v \ll 1$, that the slow roll inflation takes place.
The spectral index reads
\begin{align}
	n_s - 1 \simeq 2 \eta_{\mathcal V} - 6 \epsilon_{\mathcal V} \sim - 2 \kappa - 3 \varepsilon^2 - 48 g \frac{\varphi^2}{v^2}.
	\label{eq:ns}
\end{align}
The power spectrum is almost flat for $\kappa \sim \varepsilon^2 \lll 1$, 
while it is strongly red-tilted for $\kappa \sim \mathcal O (0.1)$.
As we see in the next section,
the PBH spectrum crucially depends on $n_s$.

While the inflaton slowly rolls down from $\varphi_\text{ini}$ to $\varphi_\text{end}$,
the scale factor grows exponentially.
The $e$-folding number of the new inflation is
\begin{align}
	N_\text{new} (\varphi)
	&\simeq \int^{\varphi}_{\varphi_\text{end}} 
	\frac{\dd \varphi}{\Mpl^2} \frac{\mathcal V}{\mathcal V'} \label{eq:efoldings_def} \\
	&\sim
	\begin{cases}
	\frac{\varphi_\ast}{\varepsilon\Mpl}
	\com{ C - \frac{\varphi}{\varphi_\ast} + \cdots } & \text{for}~~ \kappa \sim \varepsilon^2 \\[.5em]
	\frac{1}{2 \kappa}
	\ln \com{ \prn{\frac{v^2}{24 g \varphi^2}}
	\frac{\kappa + 8 g \varphi^2/v^2}{\kappa + 1/3} } & \text{for}~~ \kappa \sim \mathcal O (0.1)
	\end{cases},
	\label{eq:efoldings}
\end{align}
where $\varphi_\ast \equiv (\epsilon v^2 \Mpl/8g)^{1/3}$ is a typical field value
above which the $g \varphi^4$ term dominates over the linear term $\varepsilon \varphi$ 
in the case of $\kappa \sim \varepsilon^2$.
We have introduced an order-one constant $C\lesssim 2 \pi / 3 \sqrt{3}$.
We assume that the chaotic inflation is responsible for the large-scale perturbations 
($k \lesssim 1$-$10$ Mpc$^{-1}$) observed by Planck.
Therefore, an upper bound on the $e$-foldings number of the new inflation is imposed:
\begin{align}
	N_\text{new} \lesssim 50 + \ln \frac{v}{10^{15}\,\text{GeV}} - \frac{1}{6} \ln \frac{H_\text{new}}{H_\text{R}},
	\label{eq:upper}
\end{align}
where $H_\text{R}$ denotes the Hubble parameter at the reheating after the new inflation 
and $H_\text{new} = v^2/\sqrt{3} \Mpl$ is the Hubble parameter during the new inflation.

The new inflation generates the curvature perturbations on small scales ($k \gtrsim 1$-$10$ Mpc$^{-1}$).
It can be much larger than the observed temperature fluctuations,
which opens up a possibility to produce PBHs.
The power spectrum of the new inflation, $\mathcal P_\zeta$, is given by
\begin{align}
	\mathcal P_\zeta &\simeq \prn{ \frac{H}{2 \pi \Mpl} }^2 \frac{1}{2 \epsilon_{\mathcal V}} \\
	&\simeq \com{ \frac{1}{2 \sqrt{3} \pi}  \frac{v^2}{\Mpl^2\prn{\varepsilon + \kappa \frac{\varphi}{\Mpl }
	+ 8g \frac{\varphi^3}{v^2 \Mpl} }} }^2 \label{eq:pzeta} \\
	&\sim \prn{ \frac{1}{2 \sqrt{3} \pi} \frac{v^2}{\varepsilon \Mpl^2} }^2 ~~~\text{for} ~~~ \varphi \sim \varphi_\text{ini}.
\end{align}
One can see that a smaller $\mathbb Z_2$-breaking parameter yields
larger curvature perturbations. 
Sizable curvature perturbations, $\mathcal P_\zeta \sim \mathcal O (0.01)$, are generated for
\begin{align}
	\varepsilon = \alpha  \frac{v^2}{\Mpl^2},~~~\alpha \sim 1.
	\label{eq:z2breaking}
\end{align}
In this case, we have the following upper bound on the coupling, $g$,
so that the slow roll inflation occurs: $g \ll \Mpl^2 / v^2$.
Note that an order-one coupling of $\alpha$ is close to the minimum value
to avoid the eternal inflation
\begin{align}
	\abs{\frac{\dot \varphi }{H}} > \frac{H}{2\pi}~~ \to ~~
	\alpha  > \frac{1}{2\sqrt{3}\pi}.
\end{align}

\section{Reheating and Leptogenesis}
\label{sec: leptogen}

After the new inflation, the inflaton $\varphi$ oscillates around its potential minimum.
Its mass scale is given by
\begin{align}
	m_\varphi \sim \sqrt{\mathcal V''}|_{\varphi = \varphi_\text{min}} 
	\simeq 10^{14} \, \text{GeV} \, g^\frac{1}{4} \prn{ \frac{v}{10^{15}\, \text{GeV}} }^\frac{3}{2},
	\label{eq:mass_new}
\end{align}
and its amplitude is 
\begin{align}
	\bar\varphi \sim 
	\varphi_\text{min} \simeq
	5 \times 10^{16} \text{GeV}\, \prn{\frac{1}{g}}^\frac{1}{4} \prn{\frac{v}{10^{15}\,\text{GeV}}}^\frac{1}{2}.
\end{align}
The resonance parameter, defined in Eq.~\eqref{eq:mathieu_prm}, can be estimated as
$|q| \sim 3\times 10^{-4} \xi \, (v/10^{15} \,\text{GeV})$.
Obviously, it is much smaller than unity and thus the broad resonance does not occur.
In addition, since the cosmic expansion makes modes in the resonance band red-shifted away efficiently,
$q^2 m_\varphi / H \ll 1$, even the narrow resonance may not take place.
Although the resonance does not occur, 
some amount of Higgs fluctuations is generated perturbatively (via gravitational particle production~\cite{Ford:1986sy}),\footnote{
	An exceptional case is the conformal coupling: $\xi = 1/6$.
}
for the curvature coupling imprints the $\varphi$-oscillation.
The Higgs fluctuations are dominantly produced at the onset of $\varphi$-oscillation.
We can estimate its dispersion as~\cite{Ema:2015dka,Ema:2016hlw}
\begin{align}
	\langle h^2 \rangle \sim \frac{3}{16 \pi} \prn{\frac{6 \xi - 1}{4}}^2 \prn{\frac{m_\varphi^2 \bar \varphi^2 }{ \Mpl^2}}
	\prn{\frac{H_\text{new}}{m_\varphi}},
\end{align}
where we assume that $V \sim \varphi^2$ term dominates the $\varphi$-oscillation.
One can see that a typical Higgs field value is close to $10^{11}$ GeV for parameters of our interest.
Also, the Hubble parameter at the end of new inflation is around $10^{11}$ GeV for $v \sim 10^{15}$ GeV.
Soon after the beginning of $\varphi$-oscillation,
these scales become smaller than $h_\text{max} \sim 10^{10}$ GeV due to the cosmic expansion,
while the effective mass stabilizes the Higgs field; $\xi \overline R  \gg |\tilde\lambda| \langle h^2 \rangle$
with $\tilde \lambda \sim - 0.01$ for the best-fit values of SM parameters.

Finally, let us estimate the typical reheating temperature after the new inflation,
and discuss the compatibility with the thermal leptogenesis.\footnote{
	The observed values of the top quark and Higgs masses are consistent with thermal leptogenesis,
	for the reheating temperature can be high enough
	without destabilizing the electroweak vacuum~\cite{Espinosa:2007qp,Rose:2015lna}.
	Note here that the radiative corrections to the Higgs four-point coupling 
	form the Yukawa coupling of the right-handed neutrinos become relevant 
	only if the lightest right-handed neutrino is quite heavy, $M_1 \gtrsim 10^{13\, \sim \,14}$ GeV,
	as discussed in Ref.~\cite{EliasMiro:2011aa}.
}
Suppose that the new inflation interacts with the SM sector only through Planck-suppressed operators,
which might be natural since we consider the extraordinary flat potential [Eq.~\eqref{eq:potential_new}].
For a dimension five Planck-suppressed decay of $\varphi$,
$\Gamma_\varphi = b m_\varphi^3 / \Mpl^2$, the reheating temperature is estimated as
\begin{align}
	T_\text{R} \sim \prn{\frac{90}{\pi^2 g_\ast}}^\frac{1}{4} \sqrt{\Gamma_\varphi \Mpl}
	\sim 10^{11}\,\text{GeV}\, \prn{ \frac{b}{0.1} }^\frac{1}{2} 
	\prn{ \frac{m_\varphi}{10^{14}\,\text{GeV}} }^\frac{3}{2}.
	\label{eq:T_R}
\end{align}
The thermal leptogenesis~\cite{Fukugita:1986hr} requires the reheating temperature 
$T_\text{R} \gtrsim 10^9$ GeV~\cite{Buchmuller:2004nz}.
Interestingly, this bound is satisfied in most of the parameter space in our model.
This is because the new inflation scale, $v$, cannot be too small
since otherwise the Higgs field might fall into the true minimum,
as discussed around Eq.~\eqref{eq:bound}.

\section{Primordial Black Hole as Dark Matter}
\label{sec: pbh}

Finally, we show that PBHs are easily formed during the new inflation,
and that they can be a dominant component of DM.
A PBH  is produced, 
if an over-dense region overcomes the pressure
that prevents it from collapsing into its Schwarzchild radius.
Following the conventional arguments in Refs.~\cite{Carr:1975qj,Green:1997sz}, 
suppose a spherically symmetric over-dense region 
enters the cosmological horizon in the radiation-dominated era.
In order for such an over-dense region to overcome the pressure, 
its size, $R_c$, evaluated at a time when the over-dense region stops expanding,
should be larger than the Jeans scale, $H^{-1} / \sqrt{3}$.
The size is characterized by the density contrast at horizon crossing,
$\delta \equiv (\rho - \bar\rho ) / \bar \rho$, as $R_\text{c} \simeq \delta^{1/2} H^{-1}$.
Hence, a PBH is expected to be formed in the radiation-dominated Universe,
if the density perturbation at horizon crossing
satisfies the condition; $\delta_c < \delta $ with $\delta_c = 1/3$.
In this case, the mass of PBH is given by the horizon mass times a factor of $3^{-3/2} \simeq 0.2$~\cite{Carr:1975qj}.
There have been many attempts to refine this simple analysis 
in literature~\cite{nadezhin1978hydrodynamics,Niemeyer:1999ak,Shibata:1999zs,Musco:2004ak,Musco:2012au,Harada:2013epa,Nakama:2013ica}.
In this paper, however, we simply assume that the mass of PBH is proportional to the horizon mass
and take the traditional threshold $\delta_c = 1/3$ as a reference value.

PBH is characterized by its mass and abundance.
In the following, we briefly discuss how the new inflation parameters are related to them,
and show that PBH can be a dominant component of DM~\cite{Hawking:1971ei}.
As explained above, the mass of PBH which is formed in the radiation dominated Universe
is estimated as
\begin{align}
	M (\varphi) &= \gamma \rho \frac{4 \pi}{3} H_\text{re}^{-3}
	\simeq 4 \pi \gamma \frac{\Mpl^2}{H_\text{new}}
	\prn{\frac{H_\text{new}}{H_\text{R}}}^\frac{1}{3} e^{2N_{\text{new}}(\varphi)}, \label{eq:pbhmass} \\
	& \simeq 
	3\times 10^{25}\text{g} \,
	 \prn{\frac{\gamma}{0.2}} \prn{\frac{10^{15}\,\text{GeV}}{v}}^2 
	 \prn{ \frac{H_\text{new}}{H_\text{R}} }^\frac{1}{3}
	 e^{2N_\text{new} (\varphi) 
	 - 54},
\end{align}
where $H_\text{re}$ is the Hubble parameter at horizon re-enter of the over-dense region,
$\gamma$ is a numerical constant given by $\gamma =3^{-3/2} \simeq 0.2$ 
in the simple analytical calculation~\cite{Carr:1975qj},
and $N_\text{new} (\varphi)$ is the $e$-folding number of a mode that exits the horizon at $\varphi$ [Eq.~\eqref{eq:efoldings_def}].
Eq.~\eqref{eq:pbhmass} shows that the PBH mass strongly depends on the $e$-folding number. 
Hence, the mass can be controlled by the coupling, $g$, 
while other parameters are fixed [See Eq.~\eqref{eq:efoldings}].
The comoving momentum that exits the horizon at $\varphi$ can be expressed as
$k(\varphi) / k (\varphi_\text{ini})  = e^{N_\text{new}(\varphi_\text{ini}) - N_\text{new} (\varphi)}$.
Inverting this monotonically increasing function, we obtain $\varphi (k)$ at least numerically.
In this way, we can also write the PBH mass as a function of comoving momentum $k$: $M (k)$.
Note here that Eq.~\eqref{eq:upper} puts an upper bound on the PBH mass,
and also that the expression given in Eq.~\eqref{eq:pbhmass} is valid for
$M > 4 \pi \gamma \Mpl^2 / H_\text{R}$ because we have assumed that
PBHs are formed in the radiation dominated era.

The abundance of PBH is characterized by its mass fraction
defined as $\beta (M) \equiv \rho_\text{PBH} (M) / \rho$ at its formation
over logarithmic mass interval, $\dd \ln M$.
Assuming that the curvature perturbation follows the Gaussian statistics,
one can estimate $\beta$ as
\begin{align}
	\beta (M)
	&= 
	\int_{\delta_c} 
	\dd \delta \frac{1}{\sqrt{2 \pi \sigma^2 (M)}} e^{-\frac{\delta^2}{2 \sigma^2 (M)}}  \\
	&
	\simeq \sqrt{\frac{1}{2 \pi}} \frac{1}{\delta_c / \sigma (M)}
	e^{ - \frac{\delta_c^2}{2 \sigma^2 (M)}}.
	\label{eq:mass_frac}
\end{align}
Here $\delta_c$ is the threshold of PBH formation and
we take $\delta_c = 1/3$ as a reference value.
$\sigma (M)$ is the standard deviation of the density contrast $\delta$
associated with a PBH of mass $M$.
It is given by the coarse-grained curvature perturbation~\cite{Kawasaki:2012wr,Young:2014ana}
\begin{align}
	\sigma^2 (M) 
	&= \int \dd \log k\, W^2 (kk_{M}^{-1})\, \frac{16}{81} \prn{kk_M^{-1}}^4 \mathcal P_{\mathcal R} (k),
\end{align}
where $W$ denotes the Fourier transform of a window function smoothing over a scale $k_M^{-1}$,
$k_M$ is a comoving momentum that corresponds to a PBH of mass $M (k_M)$,
and $\mathcal P_{\mathcal R}$ denotes the power spectrum of the comoving curvature perturbation.
Note  that it coincides with that on uniform density hyper surface, 
$\mathcal P_{\mathcal R} \simeq \mathcal P_\zeta$, well outside the horizon.
In the following analysis, we take the Gaussian window function: $W (x) = e^{-x^2/2}$.
It is instructive to see its typical behavior before going into details.
A flat power spectrum, $\mathcal P_{\mathcal R} = A_0$, 
yields a flat mass variance with $\sigma^2 = (8/81)A_0$.
Also, a power law spectrum, $\mathcal P_{\mathcal R} = A_0 (k/k_0)^{n_s-1}$, 
gives a power law mass variance, 
$\sigma^2 = (8/81)\, \Gamma ((n_s + 3)/2)\, A_0 (k_M / k_0 )^{n_s - 1} $.
Therefore, as can be seen from Eqs.~\eqref{eq:ns} and \eqref{eq:pzeta},
the mass variance is almost constant for $M \gg M (\varphi_\ast)$ in the case of $\kappa \sim \varepsilon^2 \lll 1$,
while it is strongly peaked at $M \sim M(\varphi_\text{ini})$ in $\kappa \sim \mathcal O (0.1)$:
\begin{align}
	\sigma^2 (M) \sim
	\begin{cases}
		\prn{\frac{2}{243 \pi^2}}
		\prn{\frac{1}{\alpha} }^2 &\text{for}~~\kappa \sim \varepsilon^2 \\[.5em]
		\prn{\frac{2}{243 \pi^2}} 
		\prn{\frac{1}{\alpha} }^2
		\prn{\frac{M}{M (\varphi_\text{ini})}}^{\kappa} &\text{for}~~\kappa \sim \mathcal O (0.1) \\
	\end{cases},
	\label{eq:sigma2}
\end{align}
with $M \gg M(\varphi_\ast)$.

Now we are in a position to estimate the present abundance of PBHs.
The density parameter of PBHs over a logarithmic interval $\dd \ln M$ for each mass $M$
can be expressed in terms of the mass fraction $\beta (M)$:
\begin{align}
	\Omega_\text{PBH}(M) h^2 &= \left. \frac{\rho_\text{PBH} (M)}{\rho} \right|_\text{eq} \Omega_m h^2
	= \prn{\frac{T_\text{re}}{T_\text{eq}} \Omega_m h^2} \beta (M) \\
	&\simeq 
	\prn{\frac{ \beta(M) }{4\times 10^{-12}}}\,
	\prn{\frac{\gamma}{0.2}}^\frac{1}{2}
	\prn{\frac{106.75}{g_\ast (T_\text{re})}}^\frac{1}{4}
	\prn{ \frac{10^{25}\,\text{g}}{M} }^\frac{1}{2},
\end{align}
where the subscript ``eq'' indicates quantities evaluated at the matter-radiation equality,
$\Omega_m$ is the current density parameter of matter,
$T_\text{re}$ represents temperature at horizon reentry,
and $g_\ast$ is the relativistic degree of freedom in thermal plasma.
After the formation of PBH at horizon re-entry,
the energy density ratio of PBHs to radiation grows proportional to the scale factor 
because PBHs behave as matter.
The lighter PBHs that enter the horizon earlier become more abundant than the heavier one.
The additional factor, $1/M^{1/2}$, reflects this observation.
Integrating it with respect to $\dd \ln M$, we obtain the total abundance of PBH
\begin{align}
	\Omega_\text{PBH, tot} = \int \dd \ln M \, \Omega_\text{PBH} (M).
\end{align}
The density parameter of DM is $\Omega_c h^2 = 0.1198 \,(15)$~\cite{Ade:2015xua}.
Hence, PBHs can be a dominant component of DM,
if the mass fraction is around $\beta \sim \mathcal O (10^{-15 \,\sim \,-12})$ 
for $M \sim \mathcal O( 10^{20 \, \sim \, 26})$ g,
where the observational constraints are not so severe~\cite{Carr:2009jm}.
This range of mass fraction corresponds to a mass variance of $\sigma^2 \sim \mathcal O (10^{-3})$,
which indicates the $\mathbb Z_2$-breaking parameter $\varepsilon = \alpha v^2 / \Mpl^2$
with $\alpha \sim 1$, as already mentioned around Eq.~\eqref{eq:z2breaking}.
One can see that the size of the $\mathbb Z_2$-breaking parameter determines
the abundance of PBHs.

\begin{figure}[t]
\centering
	\includegraphics[width=0.49\textwidth]{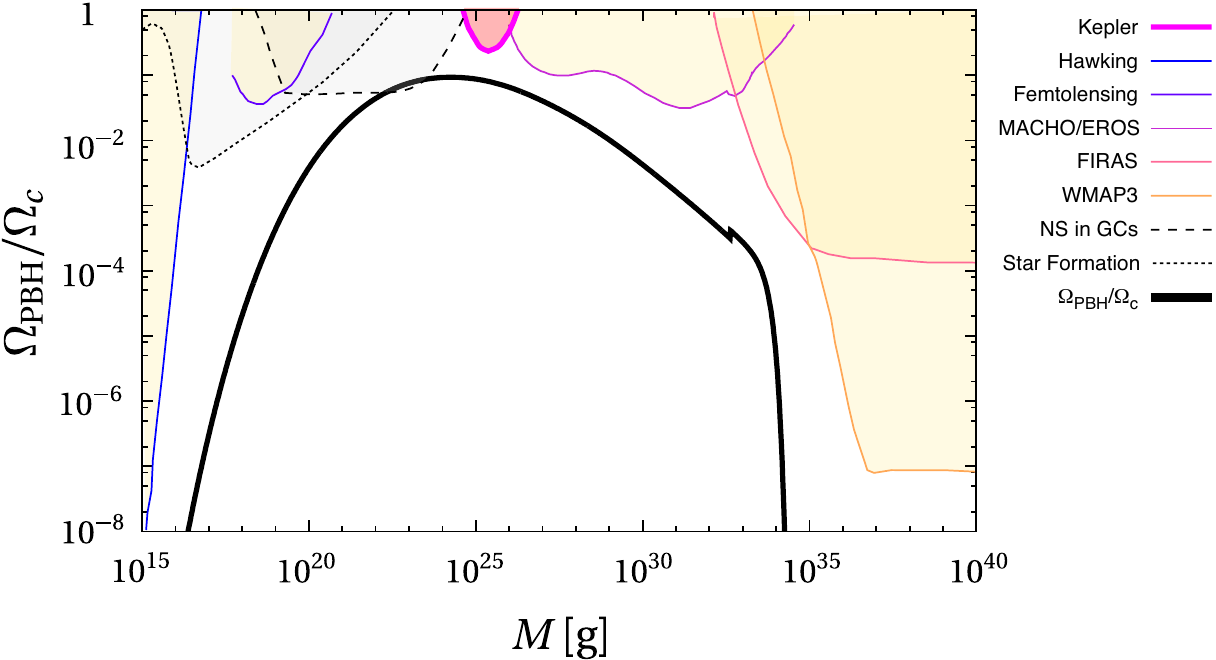} 
	\caption{\small 
	The abundance of PBHs per logarithmic interval of its mass 
	is shown in a black solid line
	in the case of $\kappa \sim \varepsilon^2$. 
	The shaded regions are excluded by observational 
	constraints~\cite{Barnacka:2012bm,Alcock:1998fx,Tisserand:2006zx,Ricotti:2007au,Griest:2013esa,Carr:2009jm,Capela:2012jz,Capela:2013yf}.
	The constraints from star formation~\cite{Capela:2012jz} and neutron star capture~\cite{Capela:2013yf} are shown in
	a gray shaded region with dotted and dashed lines.
	This is because, as claimed in Ref.~\cite{Kusenko:2013saa}, 
	the amount of DM inside globular clusters is assumed to be larger than the standard value
	and it seems to be questionable.
	We take $v = 10^{15}$ GeV, $g = 65$, and $\alpha = 0.61$.
	While its abundance for each mass is one order of magnitude smaller than that required for DM,
	the total abundance is comparable: $\Omega_\text{PBH, tot}\simeq \Omega_c$.
	Note that the pulsar timing constraints~\cite{Arzoumanian:2015liz,Lentati:2015qwp,Shannon:2015ect,Verbiest:2016vem,Janssen:2014dka} may be severe for PBHs around the solar mass 
	(See \textit{Note added}).
	Even if this is the case, we can easily evade it
	by taking a different parameter set (\textit{e.g.}~$v = 10^{15}$ GeV, $g = 82$, and $\alpha = 0.637$), 
	which does not yield such heavy PBHs but still can be a dominant component of DM.
	Details on gravitational waves (GWs) signature of our model
	will be presented elsewhere~\cite{MKT}.
	The reheating temperature after new inflation is estimated by a dimension five Planck-suppressed decay
	Eq.~\eqref{eq:T_R}.
	}
	\label{fig: flat}
\end{figure}

Note here that,
to suppress the Higgs fluctuations, which are generated during the chaotic inflaton oscillating regime,
a larger new inflation scale and $e$-folding number of the new inflation is desirable.
If these parameters are quite large,
PBHs with an interesting mass range could be generated [Eq.~\eqref{eq:pbhmass}].
Also, as indicated in Eq.~\eqref{eq:efoldings},
a larger $e$-folding number might correlate with a smaller $\mathbb Z_2$-breaking parameter
for a fixed $g \sim \mathcal O (1)$,
which opens up possibilities to produce PBHs with a sizable amount.

Finally, to be concrete, we demonstrate that PBHs can be indeed a dominant component of DM
in two benchmarks; $\kappa \sim \varepsilon^2$ and $\kappa \sim \mathcal O (0.1)$.
We will see that the resulting spectra of PBHs dramatically differ from each other.
The reheating temperature after the new inflation 
is estimated by a dimension five Planck-suppressed decay Eq.~\eqref{eq:T_R}.

In the case of $\kappa \sim \varepsilon^2$,
the mass variance is almost flat for $M (\varphi_\ast) \ll M \lesssim M (\varphi_\text{ini})$,
while it rapidly drops well below $M \ll M (\varphi_\ast)$,
as can be seen from Eqs.~\eqref{eq:sigma2} and \eqref{eq:pzeta}.
Also, Eq.~\eqref{eq:efoldings} suggests that most of the $e$-foldings number during the new inflation
is generated in the slow roll from $\varphi_\text{ini}$ to $\varphi_\ast$.
Since the PBH mass is proportional to $e^{2 N_\text{new} (\varphi)}$,
the mass fraction is flat in an extremely wide range of PBH mass,
even over ten orders of magnitude for $N_\text{new} \sim \mathcal O (10)$.
As a result, even though the abundance of PBHs per logarithmic mass interval, $\Omega_\text{PBH} (M)$, 
is one order of magnitude smaller than that required to be DM,
its total abundance can be comparable to the present DM density, $\Omega_\text{PBH, tot} \simeq \Omega_c$.
Fig.~\ref{fig: flat} shows 
the present abundance of PBHs per logarithmic mass interval 
divided by that of DM, $\Omega_\text{PBH}/\Omega_c$,
as a function of mass for one example of parameters which realize this interesting situation,
together with 
observational constraints~\cite{Barnacka:2012bm,Alcock:1998fx,Tisserand:2006zx,Ricotti:2007au,Griest:2013esa,Carr:2009jm,Capela:2012jz,Capela:2013yf}.
Note that 
the constraints from star formation and neutron star capture, shown in gray shaded regions with dotted and dashed lines,
may be questionable, 
since the assumption on the amount of DM inside globular clusters is stronger than the standard argument,
as claimed in Ref.~\cite{Kusenko:2013saa}.
Here we take $v = 10^{15}$ GeV, $g = 65$ and $\alpha = 0.61$.
One can see that a sizable amount of PBHs is produced in an extremely wide range of mass,
$10^{20}\,\text{g} \lesssim M \lesssim 10^{30}\,\text{g}$.
Though its typical abundance for each mass is one order of magnitude smaller than that of DM,
its total abundance is comparable $\Omega_\text{PBH, tot} \simeq \Omega_c$.
The spectrum has a maximum at around $M \sim 10^{24}\,\text{g}$
because the lighter PBH dominates the abundance for the flat spectrum 
with $M (\varphi_\ast) \ll M \lesssim M (\varphi_\text{ini})$
while the predicted mass fraction quickly drops for $M \ll M (\varphi_\ast)$.
	Note that the pulsar timing constraints~\cite{Arzoumanian:2015liz,Lentati:2015qwp,Shannon:2015ect,Verbiest:2016vem,Janssen:2014dka} may be severe for PBHs around the solar mass 
	because a large amount of gravitational waves (GWs) are produced as a second order effect~\cite{Saito:2008jc,Bugaev:2009zh}
	(See \textit{Note added}).
	Even if this is the case, we can easily evade it
	by taking a different parameter set (\textit{e.g.}~$v = 10^{15}$ GeV, $g = 82$ and $\alpha = 0.637$), 
	which does not yield such heavy PBHs but still can be a dominant component of DM.
	Details on GWs signature of our model
	will be presented elsewhere~\cite{MKT}.

\begin{figure}[t]
\centering
	\includegraphics[width=0.49\textwidth]{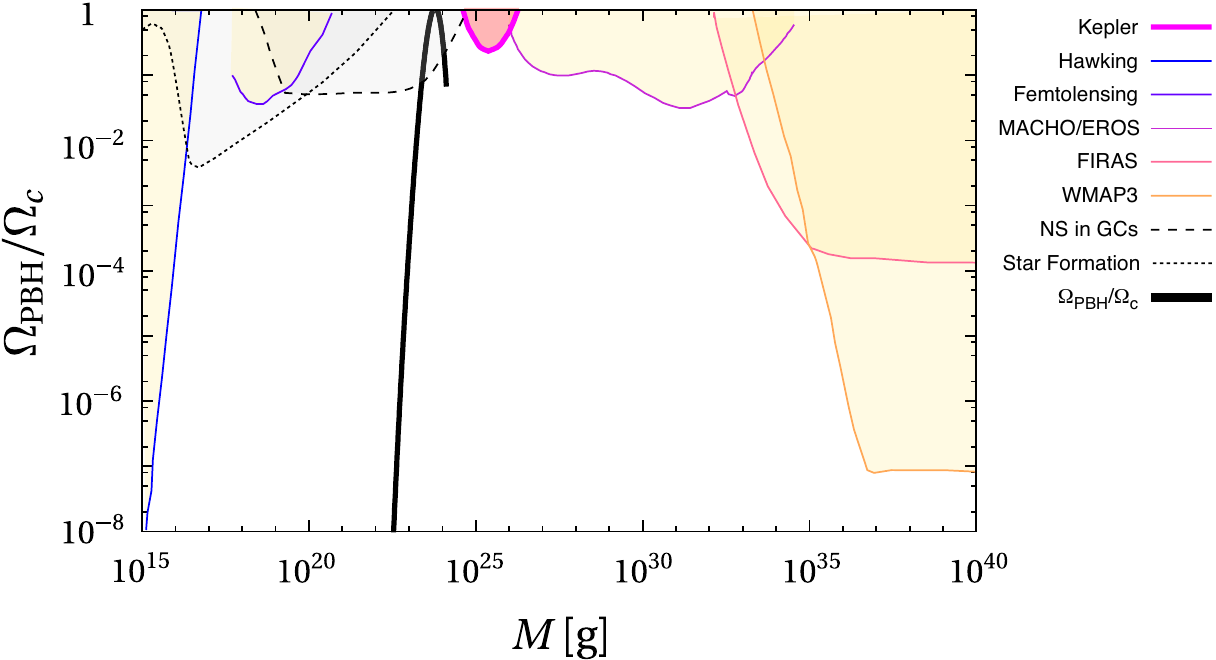}
	\caption{\small 
	The same figure as Fig.~\ref{fig: flat} in the case of $\kappa \sim \mathcal O (0.1)$.
	We take $v = 10^{15}$ GeV, $g = 0.30$, $\alpha = 0.415$, and $\kappa = 0.25$.
	The total abundance of PBHs is dominated at the sharp peak,  $M \sim 10^{23}$ g.
	The constraints from star formation and neutron star capture are questionable, for
	the assumed amount of DM inside globular clusters is larger than the standard value~\cite{Kusenko:2013saa}.
	Here, we completely neglect the constraint from neutron star.
	The reheating temperature after new inflation is estimated by a dimension five Planck-suppressed decay
	Eq.~\eqref{eq:T_R}.
}
	\label{fig: sharp}
\end{figure}

On the other hand, in the case of $\kappa \sim \mathcal O (0.1)$,
the mass variance has a sharp peak at $M \sim M(\varphi_\text{ini})$ 
as can be seen from \eqref{eq:sigma2}.
This is because the power spectrum is strongly red-tilted, $n_s - 1\simeq -2 \kappa$, due to a sizable $\kappa$.
In this case, the total abundance is dominated by this peak
and well approximated by $\Omega_\text{PBH, tot} \simeq \Omega_\text{PBH} (M (\varphi_\text{ini}))$.
Fig.~\ref{fig: sharp} shows the present abundance of PBHs per logarithmic mass interval
divided by that of DM, $\Omega_\text{PBH}/\Omega_c$,
as a function of mass, together with observational 
constraints~\cite{Barnacka:2012bm,Alcock:1998fx,Tisserand:2006zx,Ricotti:2007au,Griest:2013esa,Carr:2009jm,Capela:2012jz,Capela:2013yf}.
If we completely neglect the constraints from neutron star capture and star formation,
PBHs can be a dominant component of DM even for the sharp spectrum.
In this case,
it is clear that the abundance of PBHs is dominated by the sharp peak at around $M \sim M(\varphi_\text{ini}) \sim 10^{23}$ g.
Here we take $v = 10^{15}$ GeV, $g = 0.30$, $\alpha = 0.415$, and $\kappa = 0.25$.

\section{Conclusions}
\label{sec: conc}
In this paper, we have provided a simple cosmological solution where
chaotic inflation and the metastable electroweak vacuum are compatible.
Our scenario involves the SM plus a sizable Higgs-curvature coupling, $\xi R H^\dag H$,
and two inflations; chaotic inflation and new inflation.

The curvature coupling naturally stabilizes the Higgs field during the chaotic inflationary era,
since the Higgs field universally couples to the would-be inflatons.
Chaotic inflation generates the origin of density perturbations observed by Planck,
and also dynamically solves the initial condition problem of the new inflation.
The new inflation plays an essential role to suppress the Higgs fluctuations after inflation.
After chaotic inflation, the Higgs field acquires large fluctuations
because the curvature coupling oscillates violently.
We have shown that if the new inflation scale is sufficiently close to the chaotic one,
the Higgs field soon relaxes to the potential origin due to the Hubble-induced mass term during the new inflation
without spoiling the stability of the electroweak vacuum.
Also, we have shown that 
the production of the Higgs after the new inflation is suppressed
because the oscillation amplitude of inflaton is much smaller than the Planck scale,
contrary to the chaotic one.
Therefore, the electroweak vacuum survives during and after inflation.

One possible drawback of this model may be the coincidence of
two inflation scales.
Generally speaking, there is no a priori reason to expect that 
the new inflation scale should be close to the chaotic one,
in the context of double inflation scenarios proposed to 
solve the initial condition problem of new inflation.
However, from another viewpoint,
it yields an interesting consequence.
As explained, the electroweak vacuum becomes to be stabilized,
if the new inflation scale is sufficiently high.
Therefore, it naturally predicts the high reheating temperature.
For a dimension five Planck-suppressed decay of the inflaton, the predicted reheating temperature
is compatible with the thermal leptogenesis in most of the parameter regions.

In addition, we have shown that the new inflation naturally yields PBHs in our model.
This is because a larger $e$-folding number of the new inflation, which is desirable to relax the Higgs field,
would correlate with a smaller $\mathbb Z_2$ breaking parameter, $\varepsilon$,
for a fixed $g \sim \mathcal O (1)$.
For a smaller $\mathbb Z_2$-breaking parameter and a larger $e$-folding number,
a sizable amount of PBHs within an interesting mass range for DM  can be produced.
Note that the power spectrum of the new inflation is free from the COBE normalization,
for the chaotic one generates density perturbations observed today.
We have found interesting parameter regions where the PBHs become a dominant component of DM.
The predicted spectrum of PBHs can be either flat or sharp, depending on a model parameter.
For the flat spectrum, the PBHs are abundant in an extremely wide range of its mass, over ten orders of magnitude.
In this case, although its abundance per each mass is much smaller than that required to be 
a dominant component of DM,
its total abundance can be large enough.
On the other hand, for the sharp spectrum, the abundance is solely dominated by the peak mass.
In both cases, we have demonstrated that PBHs can be a dominant component of DM
marginally satisfying observational constraints.

\textit{Note added.\ --} GWs can be used as a useful probe for PBHs as DM
because large scalar perturbations required for the formation of PBHs generate a substantial amount of GWs
as a second-order effect~\cite{Saito:2008jc,Bugaev:2009zh}.
A typical value of the GW density parameter may be estimated as
$\Omega_\text{GW} h^2 \sim 10^{-9} ( \mathcal P_\zeta / 10^{-2})^2$
and the typical frequency is related to the PBH mass as 
$f_\text{GW} \sim 3 \times 10^{-9} \mathrm{Hz}\, (M / M_\odot)^{-1/2}$~\cite{Saito:2008jc,Bugaev:2009zh}.
Also, a merger of a PBH binary provides another source of 
GWs~\cite{Nakamura:1997sm}; the GW event observed by LIGO~\cite{Abbott:2016blz} might be the case
as discussed in~\cite{Bird:2016dcv,Clesse:2016vqa,Sasaki:2016jop}.
Those GWs can be probed 
by pulsar timing experiments at very low frequency 
$f_\text{GW} \sim 10^{-9}$--$10^{-6}\,\mathrm{Hz}$~\cite{Arzoumanian:2015liz,Lentati:2015qwp,Shannon:2015ect,Verbiest:2016vem,Janssen:2014dka},
by space-based detectors
at low frequency $f_\text{GW} \sim 10^{-5}$--$10\,\mathrm{Hz}$~\cite{Seoane:2013qna,Harry:2006fi,Seto:2001qf},
and by ground-based detectors at high frequency 
$f_\text{GW} \sim 10$--$10^2\,\mathrm{Hz}$~\cite{Punturo:2010zz,Harry:2010zz,Somiya:2011np,TheVirgo:2014hva}.

Since our model can yield a sizable amount of PBHs in an extremely wide range of its mass,
discovery of continuous spectrum of such BHs in a low-mass regime could be its signature.
Moreover, in our model, PBHs are generated from large scalar perturbations during inflation,
and hence GWs produced via those scalar fluctuations should be correlated with the abundance of PBHs.
Therefore, it is interesting to calculate the resulting GW spectrum
and discuss its signature~\cite{MKT}.

\section*{Acknowledgements}
\small\noindent
K.M.~and T.T.Y.~thank Satoshi Shirai for discussion.
This work is supported by Grant-in-Aid for Scientific Research from the Ministry of Education,
Science, Sports, and Culture (MEXT), Japan,  No.\ 15H05889 (M.K.), No.\ 25400248 (M.K.),
No.\ 26104009 (T.T.Y.), No.\ 26287039 (T.T.Y.) and No.\ 16H02176 (T.T.Y.), 
World Premier International Research Center Initiative (WPI Initiative), MEXT, Japan (M.K., K.M. and T.T.Y.),
and JSPS Research Fellowships for Young Scientists (K.M.).

\bibliographystyle{apsrev4-1}
\bibliography{higgs_pbh}
\end{document}